\documentclass{Interspeech2024}

\interspeechcameraready

\usepackage{amsmath}
\usepackage{xcolor}
\usepackage{verbatim}
\usepackage{multirow}
\usepackage{multicol}
\usepackage{comment}
\usepackage{amssymb}
\usepackage[vlined]{algorithm2e}
\usepackage{url}
\usepackage{float}
\usepackage{etoolbox}
\usepackage{adjustbox}
\usepackage{pgf}
\usepackage{soul}
\usepackage{graphicx}
\usepackage{subcaption} 
\apptocmd{\thebibliography}{\small}{}{}
\patchcmd{\thebibliography}{\leftmargin\labelwidth}{\leftmargin\labelwidth\itemsep=0pt\parsep=0pt\topsep=0pt}{}{}

\title{Are Paralinguistic Representations all that is needed for Speech Emotion Recognition?}

\name[affiliation={1}]{Orchid}{Chetia Phukan}
\name[affiliation={1}]{Gautam}{Siddharth Kashyap}
\name[affiliation={1}]{Arun}{Balaji Buduru}
\name[affiliation={1,2}]{Rajesh}{Sharma}

\address{
  $^1$IIIT-Delhi, India,\\
  $^2$University of Tartu, Estonia}
\email{orchidp@iiitd.ac.in}


\begin{document}
\maketitle

\begin{abstract}
Availability of representations from pre-trained models (PTMs) have facilitated substantial progress in speech emotion recognition (SER). 
Particularly, representations from PTM trained for paralinguistic speech processing have shown state-of-the-art (SOTA) performance for SER. However, such paralinguistic PTM representations haven't been evaluated for SER in linguistic environments other than English. Also, paralinguistic PTM representations haven't been investigated in benchmarks such as SUPERB, EMO-SUPERB, ML-SUPERB for SER. This makes it difficult to access the efficacy of paralinguistic PTM representations for SER in multiple languages. To fill this gap, we perform a comprehensive comparative study of five SOTA PTM representations. Our results shows that paralinguistic PTM (TRILLsson) representations performs the best and this performance can be attributed to its effectiveness in capturing pitch, tone and other speech characteristics more effectively than other PTM representations. 
\end{abstract}


\section{Introduction}

\label{sec:intro}
Emotions play a crucial role in human communication, influencing our behavior, decisions, and interactions. Speech emotion recognition (SER) as a task is designed to identify and understand these emotional cues conveyed through speech. By analyzing speech characteristics such as pitch, tone, intensity, and so on, 
SER models can accurately detect emotions such as happiness, sadness, anger, fear, and more. This holds immense significance across numerous domains, including human-computer interaction (HCI), healthcare, customer service, education, entertainment, as well as security. \par

Initially, research around SER mostly revolved around using traditional statistical or handcrafted features \cite{kishore2013emotion, jin2015speech}. However, with the wide-scale accessibility to pre-trained models (PTMs), the paradigm has completely shifted towards modeling SER with representations from PTMs \cite{pepino21_interspeech, yang21c_interspeech}. The wide-scale and open availability of PTMs has led to sufficient development in SER. Representations from PTMs are provided as input for downstream modeling of SER. The main reason for their wide adaptation is their performance benefit and the ability to prevent training models from scratch. The superior performance of representations from PTMs for SER  can be attributed to PTMs pre-training on diverse large-scale data, which, in return, provides meaningful representations for downstream SER. 

Previous works have exploited various PTM representations for SER such as wav2vec2 
 \cite{morais2022speech}, wavLM \cite{atmaja2022evaluating}, and so on. PTMs are trained for different tasks such as for general-purpose representation learning  \cite{chen2022wavlm}, speech recognition \cite{baevski2020wav2vec},  paralinguistic tasks \cite{shor2022universal}, etc. and with different pretext objectives as well as with different datasets. These PTMs are either trained in a single language or across multiple languages \cite{conneau2020unsupervised}. These variabilities in the PTMs, leads to variability in the downstream SER performance with representations extracted from different PTMs. As such  Morais et al. \cite{morais2022speech} have investigated representations from different PTMs for understanding the variability in the SER performance. Additionally, Phukan et al. \cite{chetiaphukan23_interspeech} have evaluated various self-supervised PTM representations alongside speaker recognition PTM representations. Interestingly, their findings indicate that speaker recognition PTM representations tend to yield superior performance compared to self-supervised PTM representations. Moreover, researchers have explored the applicability of different PTM representations for SER across multiple languages \cite{scheidwasser2022serab, singh2023decoding}. \par

Benchmarks such as SUPERB \cite{yang2021superb}, EMO-SUPERB \cite{wu2024emosuperb} further assist researchers in validating various PTM representations for SER. However, previous investigative studies as well as the benchmarks haven't explored representations from paralinguistic PTM for SER in spite of its efficacy for state-of-the-art (SOTA) performance in SER as shown by Shor et al. \cite{shor2022universal}. However, Shor et al. \cite{shor2022universal} haven't evaluated the efficacy of representations from paralinguistic PTM for SER in languages other than English. In addition, ML-SUPERB \cite{shi23g_interspeech} that evaluate PTM representations for multilingual tasks haven't included SER as a task yet. This leaves a gap for better understanding of representations from paralinguistic PTM for SER in multiple languages. So, to close this research gap, we perform a exhaustive comparative study of five PTM representations for SER consisting of representations from SOTA monolingual, multilingual, paralinguistic, as well as speaker recognition PTMs for better understanding of  paralinguistic PTM representations capability for SER. Our main contributions are as follows: 
\begin{itemize}
  \item Comparison of five PTM representations (TRILLsson, XLS-R, WavLM, Whisper, x-vector) on five benchmark datasets (CREMA-D (\textit{English}), URDU (\textit{Urdu}), BAVED (\textit{Arabic}), Emo-DB (\textit{German}), AESDD (\textit{Greek})). 
  \item Representations from paralinguistic PTM (TRILLsson) has demonstrated the topmost performance across all the datasets in comparison to representations from other PTMs which are SOTA in different benchmarks.  
  \item With TRILLsson representations, we report the best accuracy on various datasets of different languages in comparison to existing works on respective datasets. 
\end{itemize}

We are releasing the code\footnote{\url{https://github.com/orchidchetiaphukan/ParalinguisticSER}} for future works to build upon our work for effective benchmarking of SER. There are four major sections in our work. Section \ref{sec:PTME} which discusses the PTMs whose representations are under consideration for our study. Section \ref{sec:Exp} presents the datasets, modeling, and its results. Lastly, Section \ref{iiii} summarizes and concludes our study.

\section{Pre-Trained Models}\label{sec:PTME}

We use TRILLsson \cite{shor22_interspeech} as paralingustic PTM in our work. TRILLsson is built by teacher-student knowledge distillation from SOTA paralinguistic Conformer (CAP12) \cite{shor2022universal}. TRILLsson is openly available while CAP12 is not. It achieves near SOTA performance in the Non-Semantic Speech (NOSS) benchmark. NOSS consists of various non-semantic tasks such as SER, speaker recognition, synthetic audio detection, etc. AudioSet and Libri-light dataset was used for distilling TRILLsson, while CAP12 is pre-trained on YT-U. Libri-light is a 60k hours English dataset, however, YT-U may contain data in multiple languages as it is a dataset of randomly collected audios from Youtube. Non-speech-related segments were removed from the collected audios and it resulted in around 900k hours unlabeled data. We use TRILLsson\footnote{\url{https://tfhub.dev/google/nonsemantic-speech-benchmark/trillsson4/1}} available in \textit{Tensorflow Hub}. The model aggregates over time and returns a vector of 1024 dimensional size for each input audio provided. 

We use XLS-R \cite{babu22_interspeech} and Whisper \cite{radford2023robust} for multilingual PTMs. Both these PTMs are pre-trained in different manner, XLS-R in self-supervised while Whisper in weakly-supervised manner. XLS-R is pretrained on 436k hours data. Whisper is based on an encoder-decoder architecture and is trained to predict extensive volumes of audio transcriptions found on the internet. Whisper is pretrained on 680K hours encompassing 96 languages and also in multitask format. We remove the decoder and use the encoder to extract the representations. Also, we are the first work, according to best of our knowledge, to use Whisper encoder representations for multilingual SER. We use 0.3 billion parameters XLS-R\footnote{\url{https://huggingface.co/facebook/wav2vec2-xls-r-300m}} and whisper-base\footnote{\url{https://huggingface.co/openai/whisper-base}} version directly available in \textit{Hugginface}.\par

For monolingual PTM, we consider WavLM\footnote{\url{https://huggingface.co/microsoft/wavlm-base}} \cite{chen2022wavlm} because of SOTA performance in SUPERB including SER. We include x-vector\footnote{\url{https://huggingface.co/speechbrain/spkrec-xvect-voxceleb}} as speaker recognition PTM in our study as previous researchers have shown the efficacy x-vector representations for SER \cite{pappagari2020x, chetiaphukan23_interspeech}. X-vector \cite{8461375} is a time-delay neural network, trained for speaker identification in supervised manner.

The last hidden states from XLS-R, Whisper, WavLM, and x-vector are extracted and converted into 1024, 512, 768, and 512-dimensional vectors for each audio file using average pooling. Sampling is performed at a rate of 16KHz for each audio file that is supplied as input to the PTMs.

\begin{figure}[hbt!]    
\centering
      \includegraphics[width=0.42\textwidth, height=0.14\textwidth]{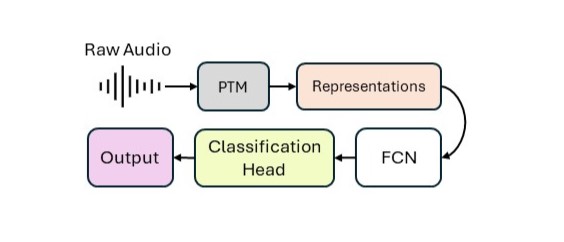}
      \caption{Modeling Approach: Fully Connected Network (FCN)}
        \label{fig:archiFCN}     
\end{figure}

\begin{figure}[hbt!]    
\centering
      \includegraphics[width=0.42\textwidth, height=0.14\textwidth]{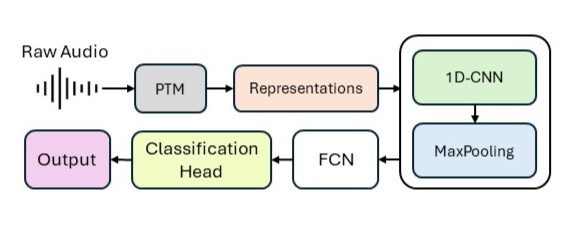}
      \caption{Modeling Approach: Convolution Neural Network (CNN)}
        \label{fig:archi}     
\end{figure}

\section{Experiments}\label{sec:Exp}

\subsection{Datasets}

\noindent\textbf{Crowd-Sourced Emotional Multimodal Actors Dataset (CREMA-D) \cite{cao2014crema}: } It is a benchmark gender-balance database in English, with 48 male and 43 female artists contributing a total of 7442 utterances. It acts as a valuable resource due to the variations in the speaker's ages and ethnicities. It consists of emotions: anger, happiness, sadness, fear, disgust, and a neutral state. The artists spoke from 12  sentences.\par
\noindent\textbf{Basic Arabic Vocal Emotions Dataset (BAVED) \cite{baved}: } It consists of three emotional states: neutral, low (tired or exhausted), and high (happiness, joy, sadness, anger). It consists of 1935 recordings from 45 male and 16 females in Arabic.\par

\noindent\textbf{URDU \cite{latif2018cross}:} It is an urdu speech-emotion dataset. In total, there are 400 spoken expressions representing four emotions: anger, happiness, sadness, and neutral. These utterances come from 27 males and 11 females. The corpus comprises of genuine and unscripted emotional segments extracted from spontaneous discussions among various guests on a television talk show.\par

\noindent\textbf{German
Emotional Speech Database (Emo-DB) \cite{burkhardt2005database}: } It is a German language dataset and consists of 535 utterances recorded from five male and five female actors. These actors were provided with a choice of ten unique scripts to deliver their lines. It contains seven emotions: anger, anxiety, boredom, disgust, happiness, neutral, and sadness.

\noindent\textbf{Acted Emotional Speech Dynamic Database (AESDD) \cite{vryzas2018speech}: } It is a Greek speech emotion dataset consisting of around 600 utterances spoken by 5 actors and comprising of five emotions: anger, disgust, fear, happiness, and sadness.

\begin{table}[ht]
\caption{Hyperparameter Details}
\centering
\label{tab:hyper}
\begin{tabular}{l|c}
\toprule
\textbf{Hyperparameter} & \textbf{Value} \\
\midrule
Number of Kernels for 1D-CNN & 32  \\  
Kernel Size for 1D-CNN & 3\\
Number of Neurons for each layer of FCN & 256, 90, 56\\
Activation Function in Intermediate Layers & ReLU\\
Training epochs & 20 \\
Optimizer & Rectified Adam \\
Learning Rate & 1e-3\\
Batch Size & 32\\
\bottomrule
\end{tabular}
\end{table}

\begin{table*}[ht]
\caption{Performance of Models trained on various PTM representations; All the scores are average of 5 folds and given in \%; F1-Score is macro average F1-Score}
\scriptsize
\centering
\begin{tabular}{l|c|c|c|c|c|c|c|c|c|c}
\toprule
\textbf{PTM} & \multicolumn{2}{c|}{\textbf{CREMA-D}} & \multicolumn{2}{c|}{\textbf{URDU}} & \multicolumn{2}{c|}{\textbf{BAVED}} & \multicolumn{2}{c|}{\textbf{Emo-DB}} & \multicolumn{2}{c}{\textbf{AESDD}} \\
\midrule
 & \textbf{Accuracy} & \textbf{F1 Score} & \textbf{Accuracy} & \textbf{F1 Score} & \textbf{Accuracy} & \textbf{F1 Score} & \textbf{Accuracy} & \textbf{F1 Score} & \textbf{Accuracy} & \textbf{F1 Score} \\
 \midrule
\multicolumn{11}{c}{\textbf{SVM}} \\
\midrule
\textbf{TRILLsson} & 79.84 & 77.66 & 93.75 & 92.75 & 85.34 & 84.80 & 92.33 & 90.93 & 89.26 & 88.89 \\
\textbf{WavLM} & 61.07 & 60.53 & 82.25 & 81.33 & 73.79 & 72.48 & 80.24 & 79.82 & 71.63 & 70.80 \\
\textbf{XLS-R} & 70.61 & 69.41 & 61.00 & 59.98 & 79.72 & 78.01 & 50.99 & 50.43 & 40.11 & 39.40 \\
\textbf{x-vector} & 63.19 & 62.83 & 84.44 & 81.33 & 80.91 & 79.44 & 85.92 & 83.53 & 74.40 & 73.30 \\
\textbf{Whisper} & 65.08 & 64.88 & 72.19 & 70.95 & 79.88 & 77.44 & 76.38 & 75.82 & 49.99 & 50.90 \\
\midrule
\multicolumn{11}{c}{\textbf{FCN}} \\
\midrule
\textbf{TRILLsson} & 81.52 & 79.84 & 95.74 & 95.80 & 86.25 & 85.99 & 93.29 & 94.2 & 91.26 & 90.89 \\
\textbf{WavLM} & 62.64 & 61.34& 85.82 & 85.45 & 77.30 & 76.79 & 81.79 & 80.6 & 73.63 & 72.84\\
\textbf{XLS-R} & 74.46 & 72.21 & 62.84 & 61.80 & 80.28 & 79.86 & 56.84 & 54.94 & 42.10 & 40.42 \\
\textbf{x-vector} & 65.88 & 65.51 & 86.34 & 82.77 & 82.29 & 81.82& 86.42 & 86.82 & 76.42 & 74.31\\
\textbf{Whisper} & 67.78 & 66.58 & 74.97 & 73.73 & 80.72 & 79.27& 80.27 & 76.01 & 52.08 & 51.92 \\
\midrule
\multicolumn{11}{c}{\textbf{CNN}} \\
\midrule
\textbf{TRILLsson} & \textbf{83.28} & \textbf{81.66} & \textbf{98.75} & \textbf{98.71} & \textbf{89.15} & \textbf{88.88} & \textbf{96.26} & \textbf{96.20} & \textbf{94.21} & \textbf{93.83} \\
\textbf{WavLM} & 65.30 & 64.99 & 86.75 & 86.32 & 78.09 & 77.57 & 82.62 & 81.47 & 74.38 & 73.58 \\
\textbf{XLS-R} & 75.22 & 74.96 & 65.50 & 64.45 & 84.13 & 83.70 & 59.44 & 51.47 & 44.63 & 42.85 \\
\textbf{x-vector} & 68.57 & 68.20 & 90.25 & 90.14 & 85.63 & 85.15 & 89.72 & 89.32 & 79.40 & 79.11 \\
\textbf{Whisper} & 70.49 & 70.29 & 77.75 & 77.51 & 83.56 & 83.11 & 77.94 & 73.79 & 55.37 & 54.47 \\
\bottomrule
\end{tabular}
\label{tableacc}
\end{table*}

\begin{table*}[ht]
\caption{Performance of Models after representations from various PTMs are projected to 512-dimension; All the scores are average of 5 folds and given in \%; F1-Score is macro average F1-Score}
\scriptsize
\centering
\begin{tabular}{l|c|c|c|c|c|c|c|c|c|c}
\toprule
\textbf{PTM} & \multicolumn{2}{c|}{\textbf{CREMA-D}} & \multicolumn{2}{c|}{\textbf{URDU}} & \multicolumn{2}{c|}{\textbf{BAVED}} & \multicolumn{2}{c|}{\textbf{Emo-DB}} & \multicolumn{2}{c}{\textbf{AESDD}} \\
\midrule
 & \textbf{Accuracy} & \textbf{F1 Score} & \textbf{Accuracy} & \textbf{F1 Score} & \textbf{Accuracy} & \textbf{F1 Score} & \textbf{Accuracy} & \textbf{F1 Score} & \textbf{Accuracy} & \textbf{F1 Score} \\
\textbf{TRILLsson} & \textbf{81.84} & \textbf{81.66} & \textbf{93.75} & \textbf{93.73} & \textbf{87.34}  & \textbf{86.80} & \textbf{95.33} & \textbf{94.93} & \textbf{90.91} & \textbf{90.81} \\ 
\textbf{WavLM} & 65.07 & 64.53 & 86.25  & 85.33 & 77.79 & 77.48 & 82.24 & 80.82 & 73.55& 71.44 \\ 
\textbf{XLS-R} & 74.61 & 74.41 & 65.00 & 63.98 & 83.72 & 83.01 & 57.01 & 51.43  & 42.56  & 40.95 \\ 
\bottomrule
\end{tabular}
\label{tableacc12}
\end{table*}

\begin{figure*}[t!]
    \centering
    \subfloat[TRILLsson\_URDU]{\includegraphics[width=0.19\linewidth, height=3.2cm, trim=2cm 2cm 2cm 2cm, clip]{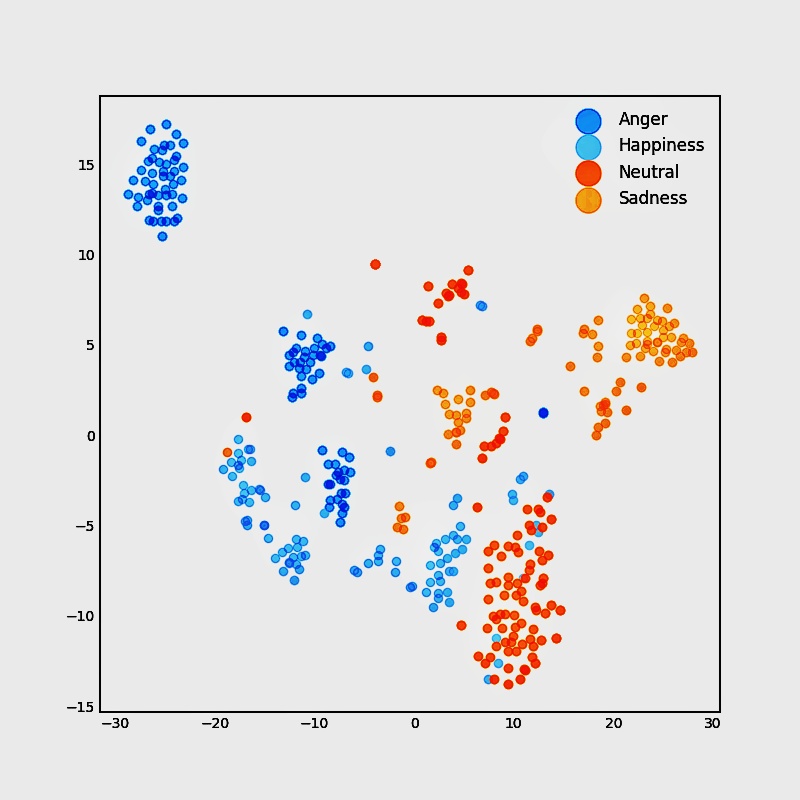}\label{uc}}
    \subfloat[XLS-R\_URDU]{\includegraphics[width=0.19\linewidth, height=3.2cm, trim=2cm 2cm 2cm 2cm, clip]{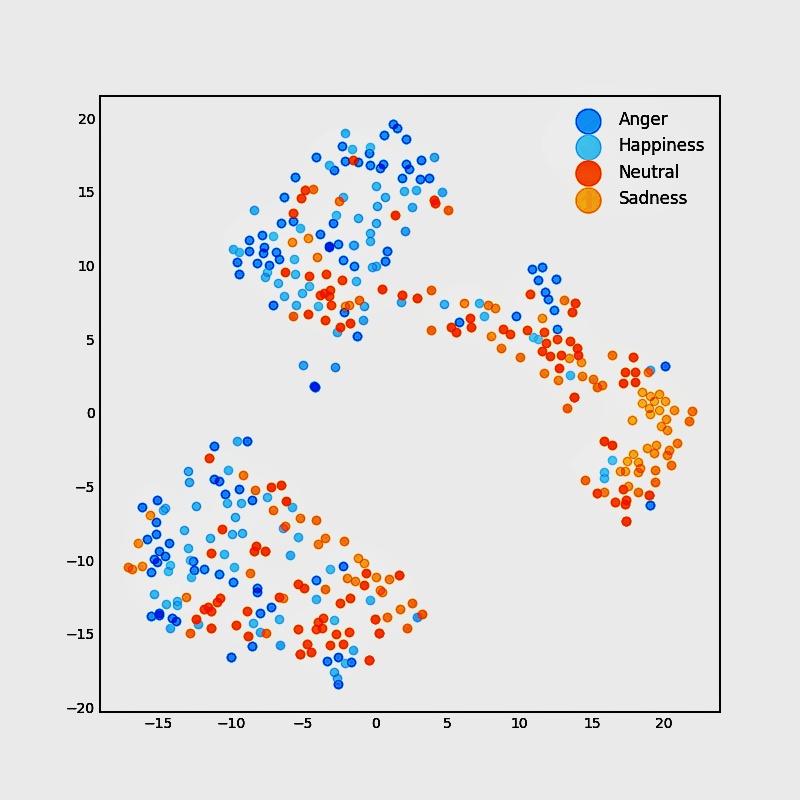}\label{uxl}}
    \subfloat[x-vector\_URDU]{\includegraphics[width=0.19\linewidth, height=3.2cm, trim=2cm 2cm 2cm 2cm, clip]{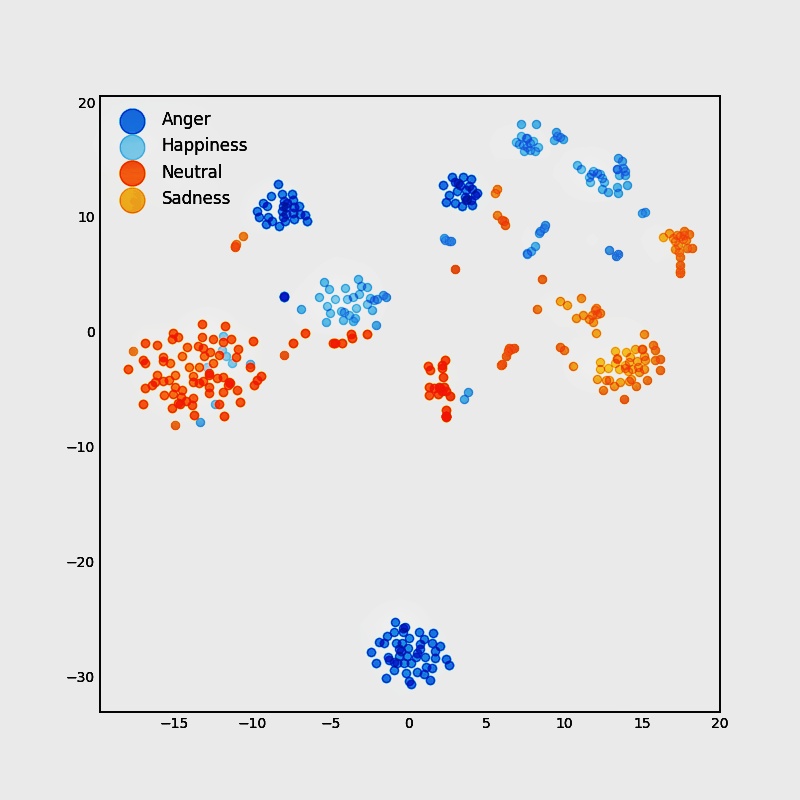}\label{uxv}}
    \subfloat[Whisper\_URDU]{\includegraphics[width=0.19\linewidth, height=3.2cm, trim=2cm 2cm 2cm 2cm, clip]{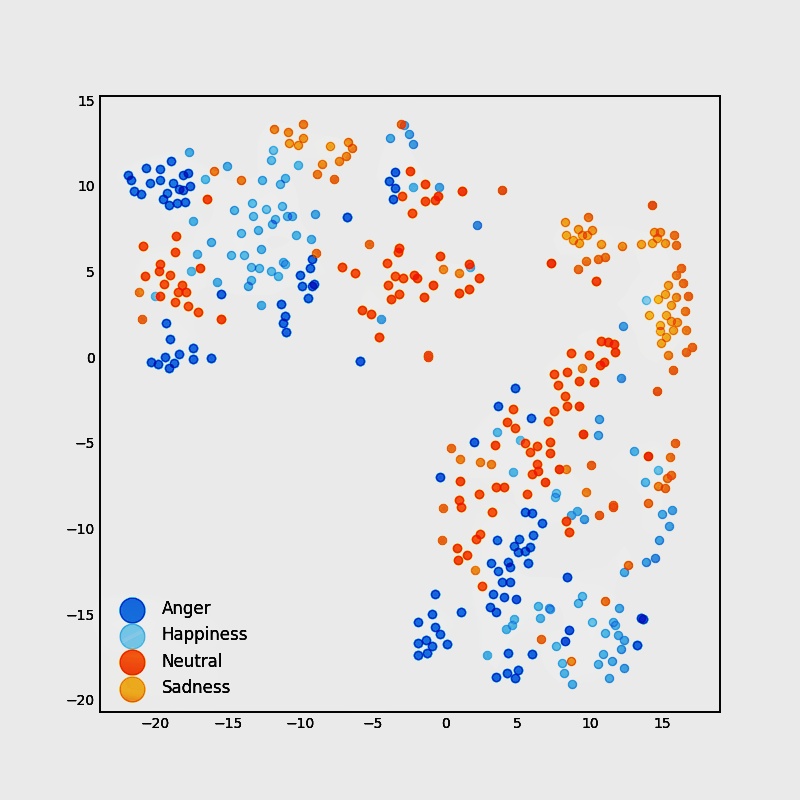}\label{uwh}}\\
    \vspace{0.2cm}
    \subfloat[TRILLsson\_Emo-DB]{\includegraphics[width=0.19\linewidth, height=3.2cm, trim=2cm 2cm 2cm 2cm, clip]{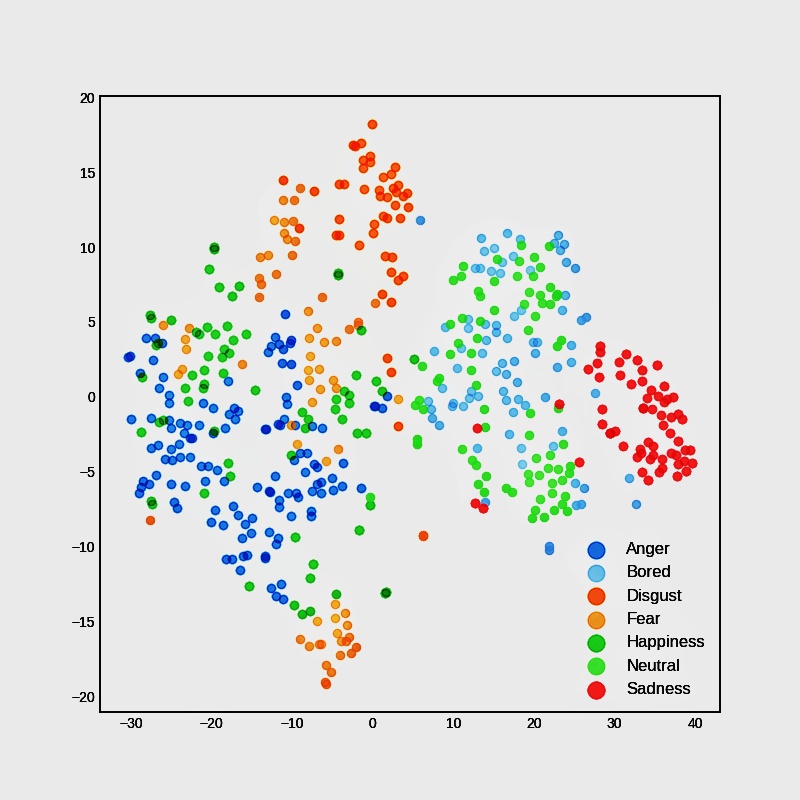}\label{ec}}
    \subfloat[XLS-R\_Emo-DB]{\includegraphics[width=0.19\linewidth, height=3.2cm, trim=2cm 2cm 2cm 2cm, clip]{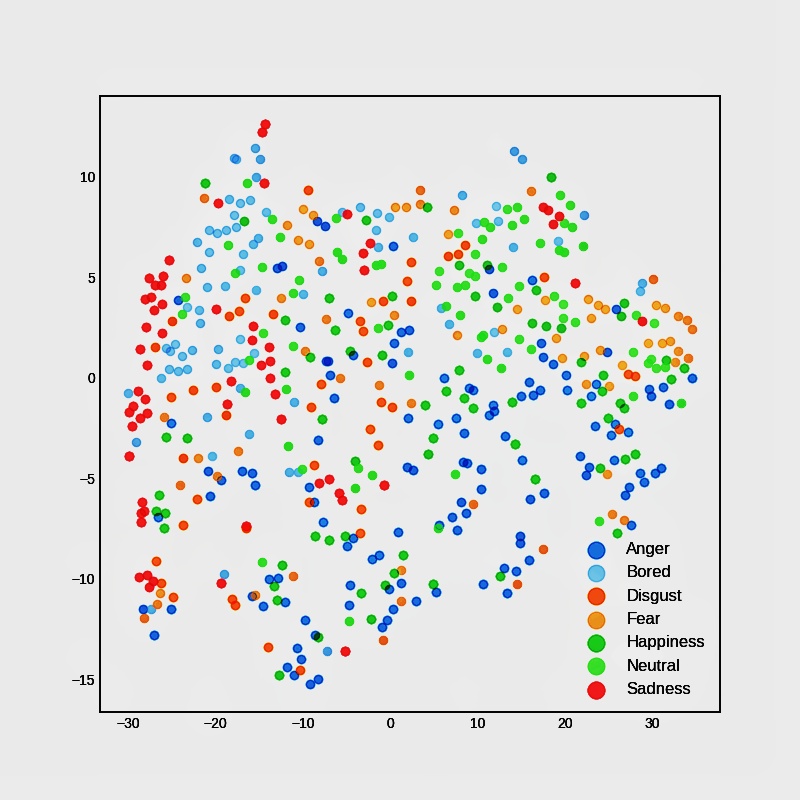}\label{exl}}
    \subfloat[x-vector\_Emo-DB]{\includegraphics[width=0.19\linewidth, height=3.2cm, trim=2cm 2cm 2cm 2cm, clip]{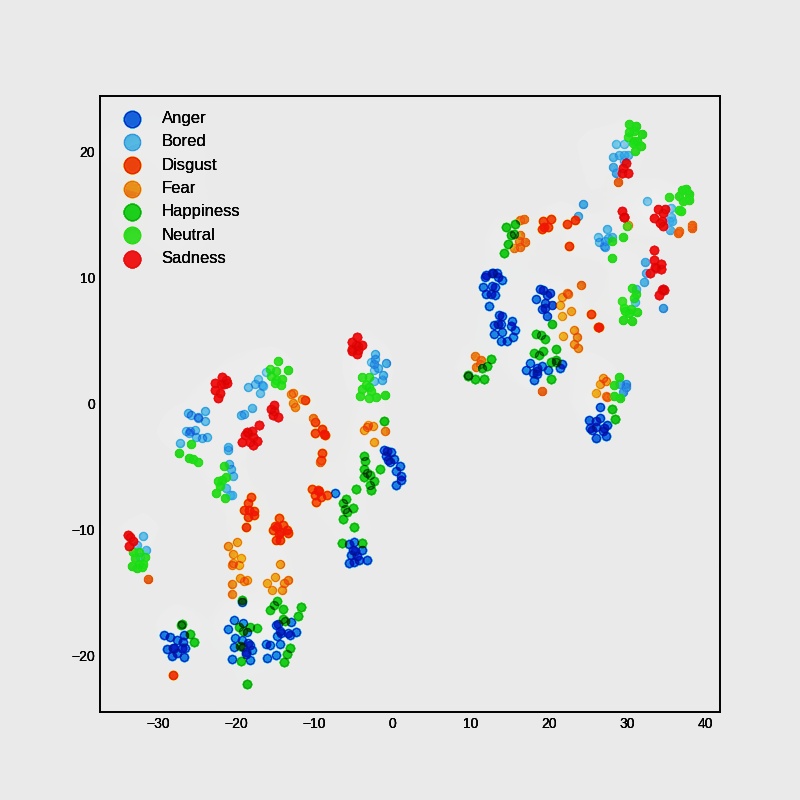}\label{exv}}
    \subfloat[Whisper\_Emo-DB]{\includegraphics[width=0.19\linewidth, height=3.2cm, trim=2cm 2cm 2cm 2cm, clip]{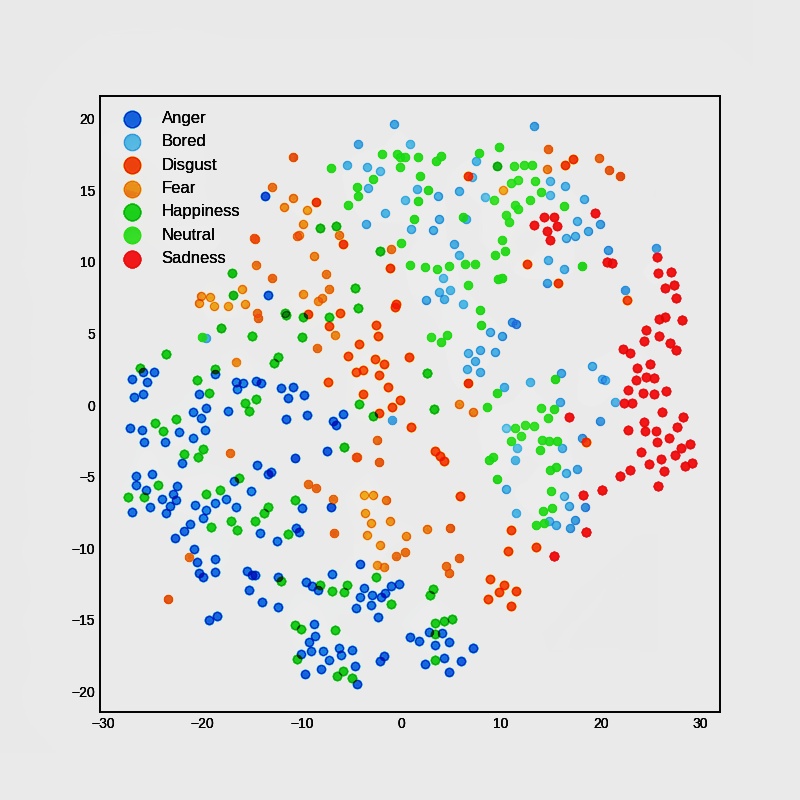}\label{ewh}}\\ 
    
   \caption{t-SNE plots of raw representations from various PTMs; Figure \ref{uc},  \ref{uxl}, \ref{uxv}, \ref{uwh} shows the t-SNE plots for URDU; Figure \ref{ec}, \ref{exl}, \ref{exv}, \ref{ewh} shows the t-SNE plots for Emo-DB}
   \label{fig:tsne} 
\end{figure*}

\subsection{Downstream Modeling}
We experiment with three downstream modeling approaches Support Vector Machine (SVM), Fully Connected Network (FCN), and CNN as these approaches have commonly used by previous studies for various related speech processing tasks \cite{kodali23_interspeech, chetiaphukan23_interspeech, mishra23_interspeech}. The modeling approaches are shown in Figure \ref{fig:archiFCN} and \ref{fig:archi}. For SVM, we kept the hyperparameters that is given by default from \textit{Scikit-Learn} library. For FCN, the extracted representations from the PTMs are directly passed to the dense layers and the number of neurons in each layer is given in Table \ref{tab:hyper}. For CNN approach, we apply 1D-CNN on top of the extracted representations from the PTM followed by a maxpooling layer (Figure \ref{fig:archi}). 1D-CNN allows extraction of further important features. The output from the maxpooling layer is flattened and passed through FCN with the same architectural settings with the FCN given in Figure \ref{fig:archiFCN}. The softmax function is used as the activation function in the classification head i.e the output layer. It outputs the probabilities that signify different emotional states. We use Cross-entropy as the loss function and \textit{Tensorflow} library for carrying out our experiments.\par

All the models are trained in a 5-fold manner with different PTM representations. Four folds are kept for training and one is for test. Details regarding the hyperparameters kept during our experiments are provided in Table \ref{tab:hyper}.

\begin{figure}[hbt!]    
\centering
      \includegraphics[width=0.45\textwidth, height=0.28\textwidth]{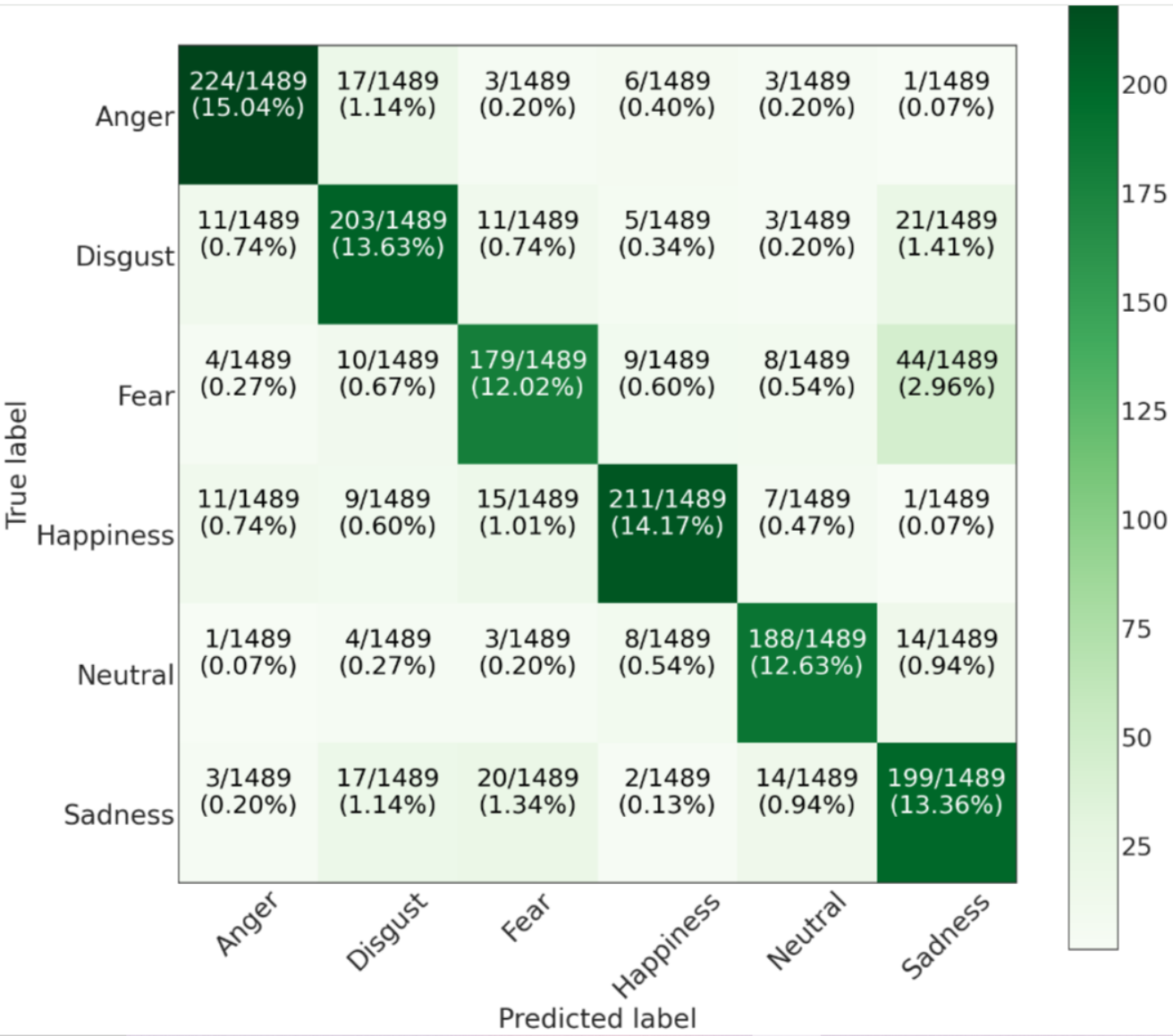}
      \caption{Confusion Matrix of CNN with TRILLsson representations for CREMA-D}
        \label{cm}     
\end{figure}

\subsection{Experimental Results}

The performance of the models trained on different PTM representations is shown in Table \ref{tableacc}. We see that models trained on TRILLsson representations performed the best across all the datasets with a sufficient margin than the other PTM representations. This demonstrates the reliability of TRILLsson representations and its capability to capture a wide range of speech characterisitcs including pitch, tone, intensity which play a significant role in influencing SER. XLS-R stands second after TRILLsson in CREMA-D, however, it fails in URDU, Emo-DB, and AESDD. WavLM representations perform comparably well despite WavLM being pre-trained only on English data. We can see mixed performance between both multilingual PTM representations, XLS-R performs better for CREMA-D and BAVED while Whisper on URDU, Emo-DB, and AESDD. This points out that the performance of the PTM representations depends on the downstream data distribution. We also plot t-SNE plots of the raw representations of various PTMs in Figure  \ref{fig:tsne}. These figures support the results obtained as better clustering across emotions is seen for TRILLsson representations compared to other PTM representations. Among the downstream models, CNN performed the best. The confusion matrix for CREMA-D for the best score i.e CNN with TRILLsson representations is shown in Figure \ref{cm}. \par

However, another important observation is the performance of x-vector representations. It is lower compared to TRILLsson representations but it is comparable to the performance of multilingual PTMs representations in certain languages and far better than them in some. This behavior could be due to the speaker recognition PTM which is able to capture certain speech characteristics present in speech that are helping for improved SER. \par

We also evaluated the PTM representations to see if dimension size has an influence on the performance. Table \ref{tableacc12} presents the results obtained. We only experiment with CNN as CNN shows the best performance amongst the downstream networks. We linear project the representations of PTMs greater than 512-dimension i.e TRILLsson, wavLM, and XLS-R to 512-dimension, which is the dimension size of Whisper and x-vector. We see there is bit drop in performance of the PTM representations when projected to a lower dimension. But, the TRILLsson representations still maintain the topmost position. x-vector and Whisper representations are not compared in Table \ref{tableacc12} as their dimension is originally 512 and its comparison is already given in Table \ref{tableacc}.



\begin{table}[htbp]
\centering
\caption{Comparison to SOTA works; CNN (TRILLsson) represents the model trained with TRILLsson representations}
\scriptsize
\begin{tabular}{l|l|c|c}
\toprule
\textbf{Dataset} & \textbf{Methods} & \textbf{Accuracy} & \textbf{F1-score} \\
\midrule
CREMA-D (\textit{English}) & SOTA \cite{shor2022universal} & \textbf{88.2} & -- \\
 & CNN (TRILLsson) & 83.28 & \textbf{81.66} \\
\midrule
URDU (\textit{Urdu}) & SOTA \cite{taj2023urdu} & 97.00 & -- \\
 & CNN (TRILLsson) & \textbf{98.75} & \textbf{98.71} \\
\midrule
BAVED (\textit{Arabic}) & SOTA \cite{alsabhan2023human} & 88.39 & -- \\
 & CNN (TRILLsson) & \textbf{89.15} & \textbf{88.88} \\
\midrule
Emo-DB (\textit{German}) & SOTA \cite{singh2023decoding} & \textbf{100} & -- \\
 & CNN (TRILLsson) & 96.26 & \textbf{96.20} \\
\midrule
AESDD (\textit{Greek}) & SOTA \cite{singh2023decoding} & 85.0 & -- \\
 & CNN (TRILLsson) & \textbf{94.21} & \textbf{93.83} \\
\midrule
emoUERJ (\textit{Portuguese}) & SOTA \cite{duret2023learning} & 97 & -- \\
 & CNN (TRILLsson) & \textbf{97.36} & \textbf{97.29} \\
\bottomrule
\end{tabular}
\label{Tablecomp}
\end{table}

\subsection{Comparison to State-of-the-Art}
We also compare our results to previous studies in Table \ref{Tablecomp}. We also experiment on an additional portugese dataset, emoUERJ \cite{germano2021emouerj}. We attain SOTA accuracy and F1-score (Macro) on 
URDU, BAVED, AESDD, and emoUERJ datasets.

\section{Conclusion}
\label{iiii}

In this work, we performed a comprehensive comparative study of five SOTA PTM representations 
for investigating the effectiveness of paralingual PTM (TRILLsson) representations for SER in multiple languages. The PTMs considered in our study are SOTA in different benchmarks. Our results shows that representations from TRILLsson performed the best among all the PTM representations and this points out its efficacy in capturing essential speech components such as pitch, tone, intensity, important for SER. Models built on TRILLson representations shows SOTA performance across various benchmark datasets. The findings of our study will be instrumental in guiding the selection of appropriate representations for SER tasks. Moreover, they draw attention to the importance of incorporating paralinguistic PTM representations into various benchmarks for SER, thus facilitating future research endeavors in this domain.

\bibliographystyle{IEEEtran}
\bibliography{main.bib}

\begin{thebibliography}{10}
\providecommand{\url}[1]{#1}
\csname url@samestyle\endcsname
\providecommand{\newblock}{\relax}
\providecommand{\bibinfo}[2]{#2}
\providecommand{\BIBentrySTDinterwordspacing}{\spaceskip=0pt\relax}
\providecommand{\BIBentryALTinterwordstretchfactor}{4}
\providecommand{\BIBentryALTinterwordspacing}{\spaceskip=\fontdimen2\font plus
\BIBentryALTinterwordstretchfactor\fontdimen3\font minus \fontdimen4\font\relax}
\providecommand{\BIBforeignlanguage}[2]{{%
\expandafter\ifx\csname l@#1\endcsname\relax
\typeout{** WARNING: IEEEtran.bst: No hyphenation pattern has been}%
\typeout{** loaded for the language `#1'. Using the pattern for}%
\typeout{** the default language instead.}%
\else
\language=\csname l@#1\endcsname
\fi
#2}}
\providecommand{\BIBdecl}{\relax}
\BIBdecl

\bibitem{kishore2013emotion}
K.~K. Kishore and P.~K. Satish, ``Emotion recognition in speech using mfcc and wavelet features,'' in \emph{2013 3rd IEEE International Advance Computing Conference (IACC)}.\hskip 1em plus 0.5em minus 0.4em\relax IEEE, 2013, pp. 842--847.

\bibitem{jin2015speech}
Q.~Jin, C.~Li, S.~Chen, and H.~Wu, ``Speech emotion recognition with acoustic and lexical features,'' in \emph{2015 IEEE international conference on acoustics, speech and signal processing (ICASSP)}.\hskip 1em plus 0.5em minus 0.4em\relax IEEE, 2015, pp. 4749--4753.

\bibitem{pepino21_interspeech}
L.~Pepino, P.~Riera, and L.~Ferrer, ``{Emotion Recognition from Speech Using wav2vec 2.0 Embeddings},'' in \emph{Proc. Interspeech 2021}, 2021, pp. 3400--3404.

\bibitem{yang21c_interspeech}
S.~wen Yang, P.-H. Chi, Y.-S. Chuang, C.-I.~J. Lai, K.~Lakhotia, Y.~Y. Lin, A.~T. Liu, J.~Shi, X.~Chang, G.-T. Lin, T.-H. Huang, W.-C. Tseng, K.~tik Lee, D.-R. Liu, Z.~Huang, S.~Dong, S.-W. Li, S.~Watanabe, A.~Mohamed, and H.~yi~Lee, ``{SUPERB: Speech Processing Universal PERformance Benchmark},'' in \emph{Proc. Interspeech 2021}, 2021, pp. 1194--1198.

\bibitem{morais2022speech}
E.~Morais, R.~Hoory, W.~Zhu, I.~Gat, M.~Damasceno, and H.~Aronowitz, ``Speech emotion recognition using self-supervised features,'' in \emph{ICASSP 2022-2022 IEEE International Conference on Acoustics, Speech and Signal Processing (ICASSP)}.\hskip 1em plus 0.5em minus 0.4em\relax IEEE, 2022, pp. 6922--6926.

\bibitem{atmaja2022evaluating}
B.~T. Atmaja and A.~Sasou, ``Evaluating self-supervised speech representations for speech emotion recognition,'' \emph{IEEE Access}, vol.~10, pp. 124\,396--124\,407, 2022.

\bibitem{chen2022wavlm}
S.~Chen, C.~Wang, Z.~Chen, Y.~Wu, S.~Liu, Z.~Chen, J.~Li, N.~Kanda, T.~Yoshioka, X.~Xiao \emph{et~al.}, ``Wavlm: Large-scale self-supervised pre-training for full stack speech processing,'' \emph{IEEE Journal of Selected Topics in Signal Processing}, vol.~16, no.~6, pp. 1505--1518, 2022.

\bibitem{baevski2020wav2vec}
A.~Baevski, Y.~Zhou, A.~Mohamed, and M.~Auli, ``wav2vec 2.0: A framework for self-supervised learning of speech representations,'' \emph{Advances in neural information processing systems}, vol.~33, pp. 12\,449--12\,460, 2020.

\bibitem{shor2022universal}
J.~Shor, A.~Jansen, W.~Han, D.~Park, and Y.~Zhang, ``Universal paralinguistic speech representations using self-supervised conformers,'' in \emph{ICASSP 2022-2022 IEEE International Conference on Acoustics, Speech and Signal Processing (ICASSP)}.\hskip 1em plus 0.5em minus 0.4em\relax IEEE, 2022, pp. 3169--3173.

\bibitem{conneau2020unsupervised}
A.~Conneau, A.~Baevski, R.~Collobert, A.~Mohamed, and M.~Auli, ``Unsupervised cross-lingual representation learning for speech recognition,'' \emph{arXiv preprint arXiv:2006.13979}, 2020.

\bibitem{chetiaphukan23_interspeech}
O.~{Chetia Phukan}, A.~{Balaji Buduru}, and R.~Sharma, ``{Transforming the Embeddings: A Lightweight Technique for Speech Emotion Recognition Tasks},'' in \emph{Proc. INTERSPEECH 2023}, 2023, pp. 1903--1907.

\bibitem{scheidwasser2022serab}
N.~Scheidwasser-Clow, M.~Kegler, P.~Beckmann, and M.~Cernak, ``Serab: A multi-lingual benchmark for speech emotion recognition,'' in \emph{ICASSP 2022-2022 IEEE International Conference on Acoustics, Speech and Signal Processing (ICASSP)}.\hskip 1em plus 0.5em minus 0.4em\relax IEEE, 2022, pp. 7697--7701.

\bibitem{singh2023decoding}
A.~Singh and A.~Gupta, ``Decoding emotions: A comprehensive multilingual study of speech models for speech emotion recognition,'' \emph{arXiv preprint arXiv:2308.08713}, 2023.

\bibitem{yang2021superb}
S.-w. Yang, P.-H. Chi, Y.-S. Chuang, C.-I.~J. Lai, K.~Lakhotia, Y.~Y. Lin, A.~T. Liu, J.~Shi, X.~Chang, G.-T. Lin \emph{et~al.}, ``Superb: Speech processing universal performance benchmark,'' \emph{arXiv preprint arXiv:2105.01051}, 2021.

\bibitem{wu2024emosuperb}
H.~Wu, H.-C. Chou, K.-W. Chang, L.~Goncalves, J.~Du, J.-S.~R. Jang, C.-C. Lee, and H.-Y. Lee, ``Emo-superb: An in-depth look at speech emotion recognition,'' 2024.

\bibitem{shi23g_interspeech}
J.~Shi, D.~Berrebbi, W.~Chen, E.-P. Hu, W.-P. Huang, H.-L. Chung, X.~Chang, S.-W. Li, A.~Mohamed, H.~yi~Lee, and S.~Watanabe, ``{ML-SUPERB: Multilingual Speech Universal PERformance Benchmark},'' in \emph{Proc. INTERSPEECH 2023}, 2023, pp. 884--888.

\bibitem{shor22_interspeech}
J.~Shor and S.~Venugopalan, ``{TRILLsson: Distilled Universal Paralinguistic Speech Representations},'' in \emph{Proc. Interspeech 2022}, 2022, pp. 356--360.

\bibitem{babu22_interspeech}
A.~Babu, C.~Wang, A.~Tjandra, K.~Lakhotia, Q.~Xu, N.~Goyal, K.~Singh, P.~{von Platen}, Y.~Saraf, J.~Pino, A.~Baevski, A.~Conneau, and M.~Auli, ``{XLS-R: Self-supervised Cross-lingual Speech Representation Learning at Scale},'' in \emph{Proc. Interspeech 2022}, 2022, pp. 2278--2282.

\bibitem{radford2023robust}
A.~Radford, J.~W. Kim, T.~Xu, G.~Brockman, C.~McLeavey, and I.~Sutskever, ``Robust speech recognition via large-scale weak supervision,'' in \emph{International Conference on Machine Learning}.\hskip 1em plus 0.5em minus 0.4em\relax PMLR, 2023, pp. 28\,492--28\,518.

\bibitem{pappagari2020x}
R.~Pappagari, T.~Wang, J.~Villalba, N.~Chen, and N.~Dehak, ``x-vectors meet emotions: A study on dependencies between emotion and speaker recognition,'' in \emph{ICASSP 2020-2020 IEEE International Conference on Acoustics, Speech and Signal Processing (ICASSP)}.\hskip 1em plus 0.5em minus 0.4em\relax IEEE, 2020, pp. 7169--7173.

\bibitem{8461375}
D.~Snyder, D.~Garcia-Romero, G.~Sell, D.~Povey, and S.~Khudanpur, ``X-vectors: Robust dnn embeddings for speaker recognition,'' in \emph{2018 IEEE International Conference on Acoustics, Speech and Signal Processing (ICASSP)}, 2018, pp. 5329--5333.

\bibitem{cao2014crema}
H.~Cao, D.~G. Cooper, M.~K. Keutmann, R.~C. Gur, A.~Nenkova, and R.~Verma, ``Crema-d: Crowd-sourced emotional multimodal actors dataset,'' \emph{IEEE transactions on affective computing}, vol.~5, no.~4, pp. 377--390, 2014.

\bibitem{baved}
A.~Aouf, ``Basic arabic vocal emotions dataset (baved),'' https://github.com/40uf411/Basic-Arabic-Vocal-Emotions-Dataset, 2019, gitHub repository.

\bibitem{latif2018cross}
S.~Latif, A.~Qayyum, M.~Usman, and J.~Qadir, ``Cross lingual speech emotion recognition: Urdu vs. western languages,'' in \emph{2018 International conference on frontiers of information technology (FIT)}.\hskip 1em plus 0.5em minus 0.4em\relax IEEE, 2018, pp. 88--93.

\bibitem{burkhardt2005database}
F.~Burkhardt, A.~Paeschke, M.~Rolfes, W.~F. Sendlmeier, B.~Weiss \emph{et~al.}, ``A database of german emotional speech.'' in \emph{Interspeech}, vol.~5, 2005, pp. 1517--1520.

\bibitem{vryzas2018speech}
N.~Vryzas, R.~Kotsakis, A.~Liatsou, C.~A. Dimoulas, and G.~Kalliris, ``Speech emotion recognition for performance interaction,'' \emph{Journal of the Audio Engineering Society}, vol.~66, no.~6, pp. 457--467, 2018.

\bibitem{kodali23_interspeech}
M.~Kodali, S.~R. Kadiri, and P.~Alku, ``{Classification of Vocal Intensity Category from Speech using the Wav2vec2 and Whisper Embeddings},'' in \emph{Proc. INTERSPEECH 2023}, 2023, pp. 4134--4138.

\bibitem{mishra23_interspeech}
J.~Mishra, J.~N. Patil, A.~Chowdhury, and M.~Prasanna, ``{End to End Spoken Language Diarization with Wav2vec Embeddings},'' in \emph{Proc. INTERSPEECH 2023}, 2023, pp. 501--505.

\bibitem{taj2023urdu}
S.~Taj, G.~M. Shaikh, S.~Hassan \emph{et~al.}, ``Urdu speech emotion recognition using speech spectral features and deep learning techniques,'' in \emph{2023 4th International Conference on Computing, Mathematics and Engineering Technologies (iCoMET)}.\hskip 1em plus 0.5em minus 0.4em\relax IEEE, 2023, pp. 1--6.

\bibitem{alsabhan2023human}
W.~Alsabhan, ``Human--computer interaction with a real-time speech emotion recognition with ensembling techniques 1d convolution neural network and attention,'' \emph{Sensors}, vol.~23, no.~3, p. 1386, 2023.

\bibitem{duret2023learning}
J.~Duret, T.~Parcollet, and Y.~Est{\`e}ve, ``Learning multilingual expressive speech representation for prosody prediction without parallel data,'' \emph{arXiv preprint arXiv:2306.17199}, 2023.

\bibitem{germano2021emouerj}
R.~B. Germano, M.~P. Tcheou, F.~da~Rocha~Henriques, and S.~P.~G. Junior, ``emouerj: an emotional speech database in portuguese,'' 2021.

\end{thebibliography}

\end{document}